\documentclass[prl,showpacs,nofootinbib,preprintnumbers,twocolumn]{revtex4}

\usepackage{graphicx}
\usepackage{amssymb}
\usepackage{pifont}

\def\lQ{\Lambda_{\rm QCD}}
\def\als{\alpha_{\rm s}}

\begin{document}

\title{Radiative decays and the nature of heavy quarkonia}
\author{Xavier \surname{Garcia i Tormo}}
\author{Joan Soto}
\affiliation{Departament d'Estructura i Constituents de la Mat\`eria, Universitat de Barcelona\\
Diagonal 647, E-08028 Barcelona, Catalonia, Spain}
\preprint{UB-ECM-PF 05/28}

\begin{abstract}

We argue that the photon spectra in radiative decays of various heavy quarkonium states provide important information on their nature.
If two of these states are in the strong coupling regime, 
we are able to produce a parameter-free model independent formula, which holds at next-to-leading order 
and includes both direct and fragmentation contributions. When the formula is checked against recent CLEO data it favors $\Upsilon (2S)$ and $\Upsilon (3S)$ in the strong coupling regime and disfavors $\Upsilon (1S)$ in it. 

\end{abstract}

\maketitle

Heavy quarkonium systems have played a major role in our understanding of QCD (see \cite{Brambilla:2004wf} for a review).
Inclusive heavy quarkonium decays to light particles involve a short distance process, in which the heavy quark and antiquark annihilate into gluons or photons, and a long distance process, which takes into account that the heavy quark and antiquark are bound in a color singlet state. Due to the asymptotic freedom of QCD the short distance process can be calculated using perturbation theory in $\als (m)$, $m$ being the heavy quark mass. It was believed for some time that in order to have a color singlet state the heavy quark and antiquark in it had to be in a color singlet state themselves. This allowed to parameterize the long distance process by a wave function at the origin (or derivatives of it). In spite of the fact that this wave function was not computable from perturbative QCD (pQCD), it canceled in suitable ratios, and hence predictions depending on pQCD only could be put forward. Nevertheless, it was soon noticed that the above framework had difficulties due to the infrared divergences  which showed up at higher orders of $\als (m)$ \cite{Barbieri:1975am}. The introduction of color octet operators in the framework of Non-relativistic QCD (NRQCD) \cite{Bodwin:1994jh} provided a rigorous way to understand, and to deal with, these divergences \cite{Bodwin:1992ye}. The price to be paid was the introduction of new non-perturbative parameters, the color octet matrix elements, which made predictions depending on pQCD only more difficult to obtain. 

Later on, it was pointed out that of the hierarchy of relevant scales in heavy quarkonium $m \gg mv \gg mv^2$, $v$ being the relative velocity, only the first inequality was exploited in NRQCD and that the velocity counting proposed in \cite{Bodwin:1994jh} did not necessarily follow from it. It was also shown how the last inequality could be exploited by constructing a further effective theory, namely Potential NRQCD (pNRQCD) \cite{Pineda:1997bj,Brambilla:1999xf}(see \cite{Brambilla:2004jw} for a review). The interplay of $\lQ$  with the scales $mv$ and $mv^2$ above dictates the degrees of freedom of pNRQCD. Two regimes have been identified : (i) the weak coupling regime, $\lQ \lesssim mv^2$, and (ii) the strong coupling regime $mv^2 \ll \lQ \lesssim mv$. In the weak coupling regime, the relevant degrees of freedom are a singlet and an octet wave function fields interacting with (perturbative) potentials and 
with 
ultrasoft gluons. The original NRQCD 
counting \cite{Bodwin:1994jh} belongs to this regime. Moreover, the use of diagrammatic and quantum mechanical methods allows to carry out explicit calculations at higher orders of $\als (mv)$ \cite{Kniehl:2002br}. In the strong coupling regime,
the essential degree of freedom is a singlet wave function field interacting with a (non-perturbative) potential. In this regime, the matrix elements of both color singlet and color octet NRQCD operators reduce to wave functions at the origin plus a few universal parameters \cite{Brambilla:2002nu}. 

Due to the fact that none of the scales involved in the hierarchies above are directly accessible experimentally, given a heavy quarkonium state, it is not obvious to which regime it must be assigned. Only the $\Upsilon (1S)$ 
appears to belong to the weak coupling regime, since weak coupling calculations in $\als (mv)$ converge reasonably well. 
The fact that the spectrum
of excitations is not Coulombic suggests that the higher excitations are not in the weak coupling regime, which can be understood from the fact that $O(\lQ)$ effects in this regime are proportional to a high power of the principal quantum number  \cite{Voloshin:1978hc}. Nevertheless, there have been claims in the literature, using renormalon-based approaches, that also $\Upsilon (2S)$ and even $\Upsilon (3S)$ can also be understood within the weak coupling regime \cite{Brambilla:2001fw}.
We argue in this letter that the photon spectra in semi-inclusive radiative decays of heavy quarkonia to light hadrons provide important information which may eventually settle this question.

The contributions to the decay width of a state $n$ can be split into direct and fragmentation, $d\Gamma_n/dz=d\Gamma^{dir}_n/dz+d\Gamma^{frag}_n/dz$.
Direct contributions are those in which the observed photon is emitted from the heavy quarks and fragmentation contributions those in which it is emitted from the decay products (light quarks). 
$z\in [0,1]$ is defined as $z=2E_\gamma /M_n$ ($M_n$ is the mass of the heavy quarkonium state), namely the fraction of the maximum energy the photon may have in the heavy quarkonium rest frame.
The approximations required to calculate the direct contributions are different in
the lower end-point region ($z\rightarrow 0$), in the central region ($z\sim
0.5$) and in the upper end-point region ($z\rightarrow 1$) of the
spectrum \cite{GarciaiTormo:2005ch} . We shall restrict our discussion to $z$ in the central region, in which no further scale is introduced beyond those
inherent of the non-relativistic system. Consequently, the photon spectrum can be
expressed in terms of matrix elements of local NRQCD operators $\mathcal{Q}$ with matching coefficients $C[\mathcal{Q}](z)$ which depend on $m$ and
$z$. 
\begin{equation}\label{dir}
\frac{d\Gamma^{dir}_n}{dz}=\sum_{\mathcal{Q}}C[\mathcal{Q}](z)\frac{\langle \mathcal{Q}\rangle_n}{m^{\delta_{\mathcal{Q}}}}
\end{equation}
$\delta_{\mathcal{Q}}$ is an integer which follows from the dimension of $\mathcal{Q}$, $\langle \mathcal{Q}\rangle_n:=\langle  V_Q (nS)\vert \mathcal{Q}\vert V_Q (nS)\rangle$ and $V_Q (nS)$ stands for a vector $S$-wave state of principal quantum number $n$.
The fragmentation contributions read \cite{Maltoni:1998nh}
\begin{eqnarray}\label{frag}
\frac{d\Gamma_n^{frag}}{dz} & =\!\!& \sum_{a = q,\bar q, g} \int_z^1\frac{dx}{x}\sum_{\mathcal{Q}}C_a[\mathcal{Q}](x)D_{a\gamma}\left(\frac{z}{x},m\right)\nonumber\\
 &:=\!\! &\sum_{\mathcal{Q}}f_\mathcal{Q}(z)\frac{\langle \mathcal{Q}\rangle_n}{m^{\delta_{\mathcal{Q}}}}
\end{eqnarray}
 $D_{a\gamma}\left(x,m\right)$ are the fragmentation functions and $C_a[\mathcal{Q}](x)$ the partonic kernels. It is important for what follows that the $f_\mathcal{Q}(z)$ are universal and do not depend on the specific bound state $n$. Due to the behavior of the fragmentation functions above, the fragmentation contributions are expected to dominate the spectrum in the lower $z$ region and to be negligible in the upper $z$ one.
In the central region, in which we will focus on,
they can be treated as a perturbation, as we will show below.

Let us first consider the weak coupling regime, for which the original NRQCD
velocity counting holds \cite{Bodwin:1994jh}. The direct contributions are
given at leading order (LO) by the $\mathcal{O}_1\left(\phantom{}^3S_1\right)$
operator; the next-to-leading order (NLO) ($v^2$ suppressed) term is given by the $\mathcal{P}_1\left(\phantom{}^3S_1\right)$ operator. The contributions of color octet operators start at order $v^4$
and are not $\als^{-1}(m)$ enhanced in the central region.
The fragmentation contributions are more difficult to organize since the importance of each term is not only fixed by the velocity counting alone but also involves the size of the fragmentation functions. It will be enough for us to restrict ourselves to the LO operators both in the singlet and octet sectors. The LO color singlet operator is $\mathcal{O}_1\left(\phantom{}^3S_1\right)$ as well.
The leading color octet contributions are $v^4$  suppressed but do have a $\als^{-1}(m)\sim 1/v^2$ enhancement with respect to the singlet ones here. They involve $\mathcal{O}_8\left(\phantom{}^3S_1\right)$, $\mathcal{O}_8\left(\phantom{}^1S_0\right)$ and $\mathcal{O}_8\left(\phantom{}^3P_0\right)$.
Then in the central region, the NRQCD expression (at the order described above) reads
\begin{eqnarray}\label{width}
\frac{d\Gamma_n}{dz} & \!\!\!=\!\!\! & \left(C_1\left[\phantom{}^3S_1\right](z)+f_{\mathcal{O}_1\left(\phantom{}^3S_1\right)}(z)\right)\frac{\langle \mathcal{O}_1(^3S_1)\rangle_n}{m^2}\nonumber\\
 & & +C_1'\left[\phantom{}^3S_1\right](z)\frac{\langle\mathcal
  {P}_1(^3S_1)\rangle_n}{m^4}+f_{\mathcal{O}_8\left(\phantom{}^3S_1\right)}(z)\nonumber \\
 & & \times\frac{\langle\mathcal{ O}_8(^3S_1)\rangle_n}{m^2}+f_{\mathcal{O}_8\left(\phantom{}^1S_0\right)}(z)\frac{\langle\mathcal{O}_8(^1S_0)\rangle_n}{m^2}\nonumber\\
 & & +f_{\mathcal{O}_8\left(\phantom{}^3P_J\right)}(z)\frac{\langle\mathcal{O}_8(^3P_0)\rangle_n}{m^4}
\end{eqnarray}
$C_1\left[\phantom{}^3S_1\right](z)$ and $C_1'\left[\phantom{}^3S_1\right](z)$ are the only short distance matching coefficients that will be eventually needed. They can be found in \cite{Brodsky:1977du} and \cite{Keung:1982jb} respectively.
If we are in the strong coupling regime and use the so called conservative counting, 
the color octet matrix elements are suppressed by $v^2$ rather than by $v^4$. Hence we should include the color octet operators in the direct contributions as well. In practise, this only amounts to the addition of $C_8$'s to the $f_{\mathcal{O}_8}$'s.
Furthermore,
$f_{\mathcal{O}_1\left(\phantom{}^3S_1\right)}(z)$, $f_{\mathcal{O}_8\left(\phantom{}^1S_0\right)}(z)$ and $f_{\mathcal{O}_8\left(\phantom{}^3P_J\right)}(z)$ are proportional to $D_{g\gamma}\left(x,m\right)$, which is small (in the central region) according to the widely accepted model \cite{Owens:1986mp}.  $f_{\mathcal{O}_8\left(\phantom{}^3S_1\right)}(z)$ is proportional to $D_{q\gamma}\left(x,m\right)$, which has been measured at LEP \cite{Buskulic:1995au}. It turns out that numerically $f_{\mathcal{O}_8\left(\phantom{}^3S_1\right)}(z)\sim C_8[\phantom{}^3S_1](z)$ in the central region. Therefore, all the LO fragmentation contributions can be treated as a perturbation. Consequently, the ratio of decay widths of two states with different principal quantum numbers is given at NLO by
\begin{eqnarray}
\frac{\displaystyle\frac{d\Gamma_n}{dz}}{\displaystyle\frac{d\Gamma_r}{dz}} & \!\!\!=\!\!\!& \frac{\langle\mathcal{O}_1(^3S_1)\rangle_n}{\langle\mathcal{O}_1(^3S_1)\rangle_r}\left(\!\!1\!+\!\frac{C_1'\left[\phantom{}^3S_1\right](z)}{C_1\left[\phantom{}^3S_1\right](z)}\frac{\mathcal{R}_{\mathcal{P}_1(\phantom{}^3S_1)}^{nr}}{m^2}\right.\nonumber\\
 & & \left.+\frac{f_{\mathcal{O}_8\left(\phantom{}^3S_1\right)}(z)}{C_1\left[\phantom{}^3S_1\right](z)}\!\mathcal{R}_{\mathcal{O}_8(\phantom{}^3S_1)}^{nr}\!\!+\!\!\frac{f_{\mathcal{O}_8\left(\phantom{}^1S_0\right)}(z)}{C_1\left[\phantom{}^3S_1\right](z)}\!\mathcal{R}_{\mathcal{O}_8(\phantom{}^1S_0)}^{nr}\right.\nonumber\\
 & & \left.+\frac{f_{\mathcal{O}_8\left(\phantom{}^3P_J\right)}(z)}{C_1\left[\phantom{}^3S_1\right](z)}\frac{\mathcal{R}_{\mathcal{O}_8(\phantom{}^3P_0)}^{nr}}{m^2}\right)
\label{nrqcd}
\end{eqnarray}
where
\begin{equation}
\mathcal{R}_{\mathcal{Q}}^{nr}=\left(\frac{\langle\mathcal{Q}\rangle_n}{\langle\mathcal{O}_1(^3S_1)\rangle_n}-\frac{\langle\mathcal{Q}\rangle_r}{\langle\mathcal{O}_1(^3S_1)\rangle_r}\right)
\label{r}
\end{equation}
Note that the $\als (m)$ corrections to the matching coefficients give rise to
negligible next-to-next-to-leading order (NNLO) contributions in the ratios above.
No further simplifications can be achieved at NLO without explicit assumptions on the counting. If the two states $n$ and $r$ are in the weak coupling regime, then
$\mathcal{R}_{\mathcal{P}_1(\phantom{}^3S_1)}^{nr}=m(E_{n}-E_{r})$ ($E_{n}$ is the binding energy) \cite{Gremm:1997dq}. 
In addition,
the ratio of matrix elements in front of the rhs of (\ref{nrqcd}) can be expressed in terms of the measured leptonic decay widths
\begin{eqnarray}\label{qucO1}
 \frac{\langle\mathcal{O}_1(^3S_1)\rangle_n}{\langle\mathcal{O}_1(^3S_1)\rangle_r} & \!\!=\!\! &  
\frac{\Gamma\left(V_Q(nS)\to e^+e^-\right)}{\Gamma\left(V_Q(rS)\to e^+e^-\right)}\nonumber\\
 & &\times\!\!\left[\!1\!-\!\frac{\mathrm{Im}g_{ee}\left(\phantom{}^3S_1\right)}{\mathrm{Im}f_{ee}\left(\phantom{}^3S_1\right)}\frac{E_{n}-E_{r}}{m}\right]
\end{eqnarray} 
$\mathrm{Im}g_{ee}$ and $\mathrm{Im}f_{ee}$ are short distance matching coefficient which may be found in \cite{Bodwin:1994jh}.
Eq.(\ref{qucO1}) and the expression for $\mathcal{R}_{\mathcal{P}_1(\phantom{}^3S_1)}^{nr}$ also hold  if both $n$ and $r$ are in the strong coupling coupling regime \cite{Brambilla:2002nu,Brambilla:2003mu}, but none of them does if one of the states is in the weak coupling regime and the other in the strong coupling regime. In the last case the NRQCD expression depends on five unknown parameters, which depend on $n$ and $r$.
If both $n$ and $r$ are in the strong coupling regime further simplifications occur. The matrix elements of the color octet NRQCD operators are proportional to the wave function at the origin times universal (bound state independent) non-perturbative parameters \cite{Brambilla:2002nu,Brambilla:2003mu}. Since  $\langle\mathcal{O}_1(^3S_1)\rangle_n$ is also proportional to the wave function at the origin, the latter cancels in the ratios involved in (\ref{r}).
Hence, 
$\mathcal{R}_{\mathcal{Q}}^{nr}=0$ for the octet operators appearing in (\ref{nrqcd}). Then, the pNRQCD expression for the ratio of decay widths reads
\begin{eqnarray}\label{scsc}
\frac{\displaystyle\frac{d\Gamma_n}{dz}}{\displaystyle\frac{d\Gamma_r}{dz}} & \!\!=\!\! & 
\frac{\langle\mathcal{O}_1(^3S_1)\rangle_n}{\langle\mathcal{O}_1(^3S_1)\rangle_r}\nonumber\\
 & & \times\!\!\left(1+\frac{C_1'\left[\phantom{}^3S_1\right](z)}{C_1\left[\phantom{}^3S_1\right](z)}\frac{1}{m}\left(E_{n}-E_{r}\right)\right)
\end{eqnarray}
Therefore, in the strong coupling regime 
we can predict
, using pNRQCD, 
the ratio of photon spectra at NLO (in the $v^2$, $(\lQ /m )^2$ \cite{Brambilla:2002nu} and $\als (\sqrt{m\lQ})\times \sqrt{ \lQ /m} $ \cite{Brambilla:2003mu} expansion),
which is the main result of this letter. 
On the other hand, if one of the states $n$ is in the weak coupling regime of pNRQCD, $\mathcal{R}_{\mathcal{Q}}^{nr}$ will have a non-trivial dependence on the principal quantum number $n$  and hence it is not expected to vanish.
Therefore, expression (\ref{scsc}) provides invaluable help for identifying the 
nature of heavy quarkonium states. If the two states are in the strong coupling regime, the ratio must follow the formula (\ref{scsc}); on the other hand, if (at least) one of the states is in the weak coupling regime the ratio is expected to deviate from (\ref{scsc}), 
and should follow the general formula (\ref{nrqcd}). We illustrate the expected deviations in the plots (dashed curves) by assigning to the unknown $\cal R$s in (\ref{nrqcd}) the value $v^4$ ($v^2 \sim 0.1$), according to the original NRQCD velocity scaling.

\begin{figure}
\centering
\includegraphics[width=7.5cm]{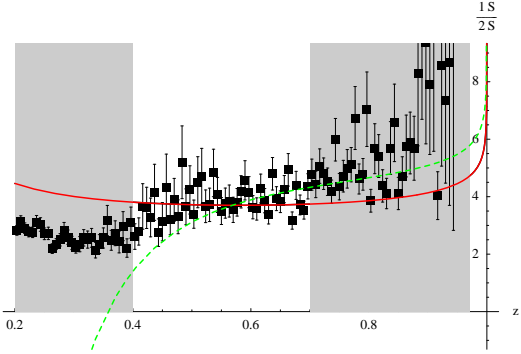}
\caption{Ratio of the $\Upsilon (1S)$ and $\Upsilon (2S)$ photon spectra. The points are obtained from the CLEO data \cite{Besson:2005jv}. The solid line is eq.(\ref{scsc}) (overall normalization fitted), the dashed line is the estimate of (\ref{nrqcd}) (see text). Agreement between the solid curve and the points in the central (unshaded) region would indicate that the two states are in the strong coupling regime.}
\label{fig1s2s}
\end{figure}
\begin{figure}
\centering
\includegraphics[width=7.5cm]{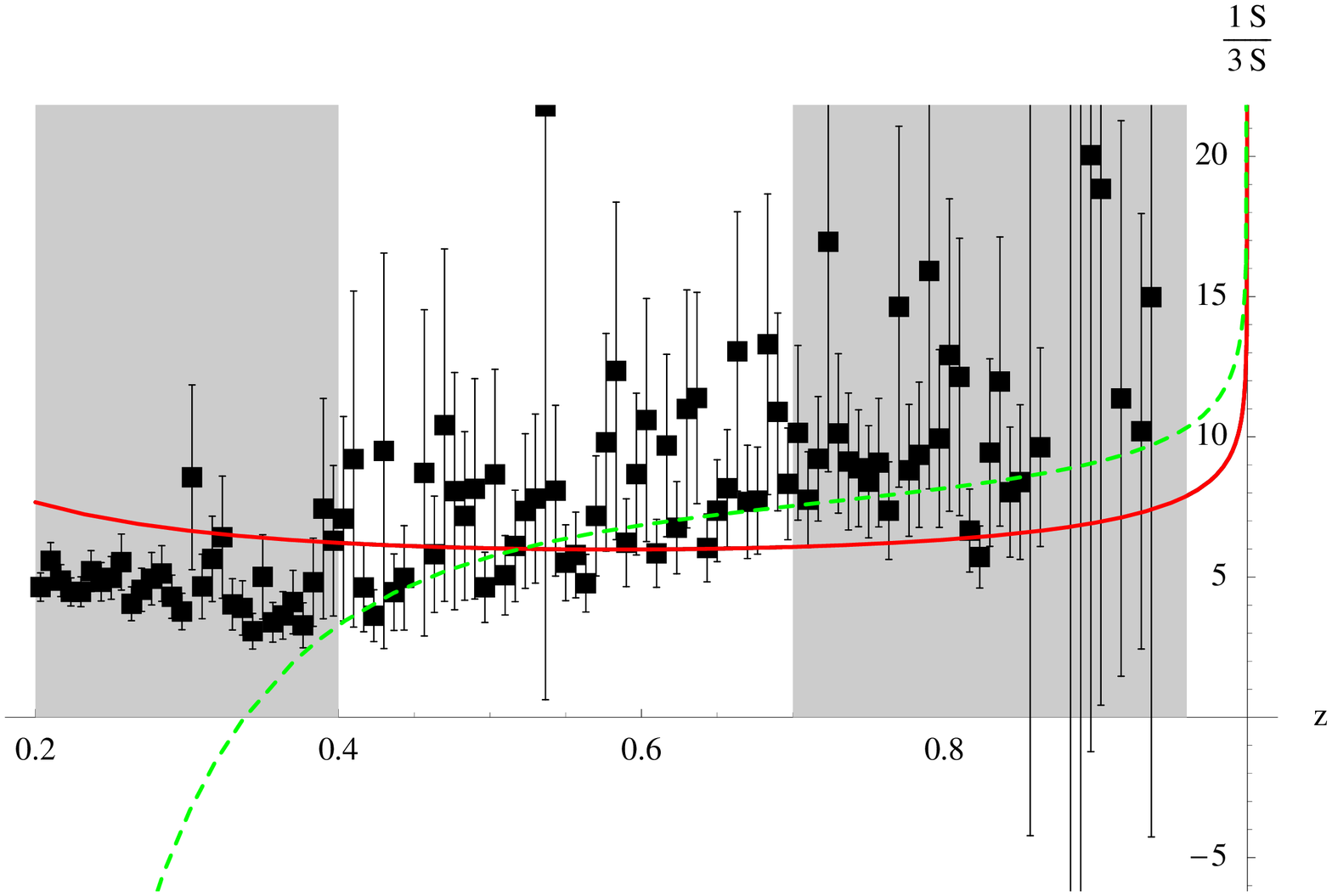}
\caption{Same as fig. 1 for 
$\Upsilon (1S)$ and $\Upsilon (3S)$ 
.} 
\label{fig1s3s}
\end{figure}
\begin{figure}
\centering
\includegraphics[width=7.5cm]{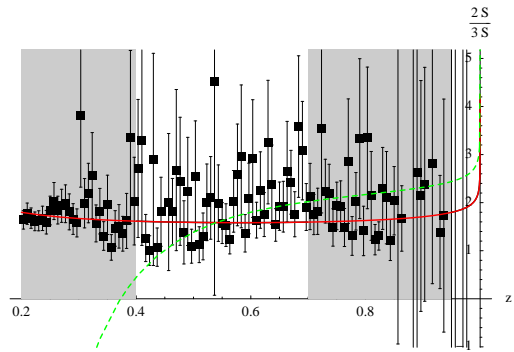}
\caption{Same as fig. 1 for 
$\Upsilon (2S)$ and $\Upsilon (3S)$ 
.} 
\label{fig2s3s}
\end{figure}

Recently CLEO \cite{Besson:2005jv} has measured the photon spectrum for the $\Upsilon (1S)$ radiative decay (with very good precision) and also (for the first time) the spectra for the $\Upsilon (2S)$ and $\Upsilon (3S)$ radiative decays. We will use this data to check our predictions. In order to do the comparison we use the following procedure. First we efficiency correct the data (using the efficiencies modeled by CLEO). Then we perform the ratios $1S/2S$, $1S/3S$ and $2S/3S$ (we add the errors of the different spectra in quadrature). Now we want to discern which of these ratios follow eq.(\ref{scsc}) and which ones deviate from it; to do that we fit eq.(\ref{scsc}) to each of the ratios leaving only the overall normalization as a free parameter (the experimental normalization is unknown). The fits are done in the central region, that is $z\in[0.4,0.7]$, where eq.(\ref{scsc}) holds. A good (bad) $\chi^2$ obtained from the fit will indicate that the ratio 
does (not) follow the shape dictated by eq.(\ref{scsc}).
In figures \ref{fig1s2s}, \ref{fig1s3s} and \ref{fig2s3s} we plot the ratios $1S/2S$, $1S/3S$ and $2S/3S$ (respectively) together with eq.(\ref{scsc}) and the estimate of (\ref{nrqcd}) mentioned above (overall normalizations fitted for all curves, the number of d.o.f. is then $45$). The figures show the spectra for $z\in[0.2,1]$ for an easier visualization but remember that we are focusing in the central $z$ region, denoted by the unshaded region in the plots. The theoretical errors due to higher orders in $\als (m)$ and in the expansions below (\ref{scsc}) are negligible with respect to the experimental ones. For the $1S/2S$ ratio we obtain 
a $\chi^2/\mathrm{d.o.f.}\vert_{1S/2S}\sim 1.2$, which corresponds to an $18\%$ CL.
The errors for the $\Upsilon (3S)$ photon spectrum are considerably larger than those of the other two states, this causes the ratios involving the $3S$ state to be less conclusive than the other one. In any case we obtain 
$\chi^2/\mathrm{d.o.f.}\vert_{1S/3S}\sim 0.9$, 
which corresponds to a $68\%$ CL,
and  $\chi^2/\mathrm{d.o.f.}\vert_{2S/3S}\sim 0.75$ 
which corresponds to an $89\%$ CL. 
Hence, the data disfavors $\Upsilon (1S)$ in the strong coupling regime but is consistent with $\Upsilon (2S)$ and $\Upsilon (3S)$ in it. For completeness we also quote the numbers corresponding to the estimate of eq. (\ref{nrqcd}) (dashed lines):  $\chi^2/\mathrm{d.o.f.}\vert_{1S/2S}\sim 1.93$ ($.02\%$ CL), $\chi^2/\mathrm{d.o.f.}\vert_{1S/3S}\sim .56$ ($99\%$ CL) and $\chi^2/\mathrm{d.o.f.}\vert_{2S/3S}\sim .66$ ($96\%$ CL). Recall that these curves are only estimates to illustrate what the differences from eq. (\ref{scsc}) to eq. (\ref{nrqcd}) may be and do not intend to best fit data.

In summary, using pNRQCD we have worked out a model-independent formula which involves the photon spectra of two heavy quarkonium states and holds at NLO in the strong coupling regime.
When this formula is applied to the Upsilon system, current data indicate that the $\Upsilon (2S)$ and the $\Upsilon (3S)$  are consistent as states in the strong coupling regime whereas the $\Upsilon (1S)$ in this regime is disfavor. A decrease of the current experimental errors for $\Upsilon (2S)$ and, specially, for the $\Upsilon (3S)$ 
is necessary to confirm this indication. This is important, since it would validate the use of the formulas in \cite{Brambilla:2002nu,Brambilla:2003mu}, and others which may be derived in the future under the same assumptions, not only for the $\Upsilon (2S)$ and $\Upsilon (3S)$ but also for the $\chi_b (2P)$s, since their masses lie in between, as well as for their pseudoscalar partners. 


\begin{acknowledgments} 

We thank D. Besson and S. Henderson for providing us with the data of ref.\cite{Besson:2005jv}, and Ll. Garrido for useful discussions. We acknowledge financial support from 
the MEC (Spain) grant CYT FPA 2004-04582-C02-01, the CIRIT (Catalonia) grant 2005SGR00564 and the network Euridice (EU) HPRN-CT2002-00311. X.G.T. acknowledges financial support from the DURSI of the Generalitat de Catalunya and the Fons Social Europeu.

\end{acknowledgments}

\end{document}